\documentclass[bm,aps,showpacs,amsfonts,amssymb,preprint,nofootinbib]{revtex4}

\usepackage{graphicx}
\usepackage{natbib}
\usepackage{amsmath}

\usepackage{color}

\font\mybb=msbm10 at 12pt

\def\bb#1{\hbox{\mybb#1}}

\def\QQ {\bb{Q}}

\def\PP {\bb{P}}

\def\FF {\bb{F}}

\def\CC{{\bf C}}

\newcommand\beqa{\begin{eqnarray}}
\newcommand\eeqa{\end{eqnarray}}
\newcommand\n{\nonumber\\}

\begin{document}

{~}

\title{Looijenga's weighted projective space, Tate's algorithm and 
Mordell-Weil Lattice in F-theory and heterotic string theory }

\vspace{2cm}
\author{Shun'ya Mizoguchi\footnote[1]{E-mail:mizoguch@post.kek.jp} and Taro Tani\footnote[2]{E-mail:tani@kurume-nct.ac.jp}}

\vspace{2cm}

\affiliation{\footnotemark[1]Theory Center, Institute of Particle and Nuclear Studies,
KEK\\Tsukuba, Ibaraki, 305-0801, Japan 
}
\affiliation{\footnotemark[1]SOKENDAI (The Graduate University for Advanced Studies)\\
Tsukuba, Ibaraki, 305-0801, Japan 
}

\affiliation{\footnotemark[2]Kurume National College of Technology, \\
Kurume, Fukuoka, 830-8555, Japan 
}
\begin{abstract} 
It is now well known that the moduli space of a vector bundle for 
heterotic string compactifications to four dimensions   
is parameterized by a set of sections of a weighted projective space 
bundle of a particular kind, known as Looijenga's weighted projective space bundle.  
We show that the requisite weighted projective spaces 
and the Weierstrass equations describing the spectral covers for 
gauge groups 
$E_N$ $(N=4,\cdots,8)$ and  $SU(n+1)$ $(n=1,2,3)$ 
can be obtained systematically by a series 
of blowing-up procedures according to Tate's algorithm, 
thereby the sections of correct line bundles claimed to arise by Looijenga's  
theorem can be automatically obtained. 
They are nothing but 
the four-dimensional analogue 
of 
the set of independent polynomials  in 
the six-dimensional F-theory  
parameterizing the complex structure, which is further confirmed  
in the constructions of $D_4$, $A_5$, $D_6$, $E_3$ 
and $SU(2)\times SU(2)$ bundles.
%
We also explain why we can obtain 
them in this way by using the structure theorem of the 
Mordell-Weil lattice, 
which is also useful for understanding the relation between 
the singularity and the occurrence of chiral matter in F-theory.

\end{abstract}

\preprint{KEK-TH 1909}
\pacs{
}
\date{July 18, 2016}

\maketitle

\section{Introduction}
It is now a well-known fact that holomorphic vector bundles on an 
elliptically fibered Calabi-Yau, needed for 
heterotic string compactifications to four dimensions, are 
constructed by considering spectral covers \cite{FMW}.
A spectral cover is basically a ramified $n$-fold cover (for 
an $SU(n)$ bundle) of the base of the elliptic Calabi-Yau, 
representing the Wilson lines in each elliptic fiber as points 
on the elliptic fiber identified with its dual. One then introduces 
a twisting line bundle over the base whose first Chern class 
(the $\eta$ class) is related to the number of instantons of the bundle.
Once one has a spectral surface and a line bundle over it, 
one can construct a vector bundle via the Fourier-Mukai 
transform using the Poincare bundle, up to a so-called $\gamma$ class 
corresponding to the $G$-flux in the F-theory dual. For more detail of 
the spectral cover construction, see {\em e.g.} 
\cite{FMW,Curio,DiaconescuIonesei,DonagiWijnholt}.

It is also now well known that the moduli space of the vector bundle 
is parameterized by a set of sections of a weighted projective space 
bundle of a particular kind, known as Looijenga's weighted projective space bundle.  
Some time ago, for $E_8$, $E_7$ and $E_6$ 
bundles and other lower-rank ones 
Looijenga's theorem was confirmed  \cite{FMW}  (except for some 
subtlety in $E_8$) by explicitly constructing spectral covers by 
using del Pezzo surfaces.  Although this approach was enough 
to reveal the validity of the miraculous nature of Looijenga's theorem,  
the constructions of the bundles were done case by case and appear 
to be independent and unrelated with each other. 
In this paper we will show that the requisite weighted projective spaces 
and the Weierstrass equations describing the spectral covers for 
$E_8$ through $A_1$ bundles can be 
obtained systematically by a series of blowing up procedures according 
to the well-known Tate's algorithm, thereby the sections of correct 
line bundles claimed to arise by the theorem can be automatically 
obtained. We will also explain why we can obtain 
them in this way by using the structure theorem of the 
Mordell-Weil lattice \cite{OguisoShioda}.

We will also show that the structure theorem of the 
Mordell-Weil lattice is useful for understanding the relation between 
the singularity and the occurrence of chiral matter in F-theory. (This new role 
of the Mordell-Weil lattice in F-theory was already briefly discussed 
in \cite{MizoguchiTani}.) 
In the literature the relation between sections and the appearance of chiral matter
is somewhat indirect. That is, on the heterotic side 
one considers a vector bundle on a spectral cover 
and computes the cohomology by means of the Leray spectral sequence to find that 
the chiral matter is localized where one or some of the ``matter curves'' representing 
the Wilson lines go(es) to infinity (zero in the addition rule of the $\wp$ function). 
On the F-theory side, the del Pezzo surface (or rational elliptic surface) itself 
in which the spectral cover is defined becomes the fiber with an appropriate 
twist corresponding to the weighted projective space bundle of Looijenga, 
and matter arise where the singularity is enhanced \cite{MorrisonVafa,BIKMSV,KatzVafa}
along the 7-brane. We will show that the structure of the 
Mordell-Weil lattice ensures the compatibility of these two pictures of chiral 
matter generations.

The plan of this paper is as follows. In section 2, we start from the same 
degree six equation in the weighted projective space 
${\rm W}\PP^3_{(1,1,2,3)}$ given in \cite{FMW} for $E_8$ bundles and review 
the construction of the spectral cover. Then we tune some of the sections 
to be in a special form so that the $dP_9$ develops a singularity. It turns out 
that, by blowing up the singularity, we are automatically led to the equation 
for $E_7$ bundles discussed in \cite{FMW}, where the relevant sections 
are precisely the ones constituting the correct weighted projective space 
of Looijenga. Repeating a similar procedure, we will find a series of 
spectral covers for the vector bundles from $E_7$ through $A_1$. 
In section 3  
we will see that the sections parameterizing a Looijenga's 
weighted projective space are nothing but 
the four-dimensional analogue 
of 
the set of independent polynomials  in 
the six-dimensional F-theory  
parameterizing the complex structure of the elliptic manifold 
with a singularity orthogonal  to the gauge group of 
the vector bundle in the whole $E_8$. This fact is further confirmed 
in the constructions of $D_4$, $A_5$, $D_6$, $E_3$ and $SU(2)\times SU(2)$ bundles.
In section 4 we discuss why this is possible by introducing the structure 
theorem of the Mordell-Weil lattice. The final section is devoted to conclusions.

\section{Tate's algorithm and Looijenga's weighted projective space}
\subsection{$E_8$ bundles: the generic case}
We start with the construction of $E_8$ bundles, following \cite{FMW}. 
As pointed out there, it is well known that $E_8$ bundles have exceptional 
features
but the construction is important because it is the 
starting point of all the constructions of the vector bundles for other 
gauge groups of lower ranks.

Let us first consider a generic degree six equation in 
${\rm W}\PP^3_{(1,1,2,3)}$ \cite{FMW} 
with homogeneous coordinates
$(u,v,X,Y)\sim(\lambda u,\lambda v,\lambda^2 X,\lambda^3 Y)$ $(\lambda\in \CC)$:
\beqa
0&=&Y^2+X^3-\frac{g_2}4 X v^4-\frac{g_3}4 v^6\n
&&+(\beta_4 u^4+\beta_3 u^3 v+ \beta_2 u^2 v^2 +\beta_1 u v^3)X\n
&&+(\alpha_6 u^6 + \alpha_5 u^5 v +\alpha_4 u^4 v^2 +\alpha_3 u^3 v^3 +\alpha_1 u v^5).
\label{E8W1123}
\eeqa
In a patch $u\neq 0$, we define affine coordinates $(z,x,y)$
by $(u,v,X,Y)\sim (1,\frac v u,\frac X{u^2},\frac Y{u^3})
\equiv(1,z,x,y)$. Then we have 
\beqa
0&=& y^2+ x^3-\frac{g_2}4 x z^4-\frac{g_3}4 z^6\n
&&+(\beta_4 +\beta_3 z+ \beta_2 z^2 +\beta_1 z^3) x\n
&&+(\alpha_6 + \alpha_5 z +\alpha_4 z^2 +\alpha_3 z^3 +\alpha_1 z^5).
\eeqa
(\ref{E8W1123}) is a $dP_8$, and by blowing up at $u=v=0$ it becomes a 
rational elliptic surface $dP_9$ with section 
\cite{FMW}. Then it can be viewed as an elliptic fibration over $\PP^1$ whose 
coordinates are $(u:v)$, the affine coordinate being $z$ in the affine patch $u\neq 0$.

To serve as a part of compactifications of F-theory \cite{Ftheory} to lower 
dimensions,  this $\PP^1$ must be further fibered over some base space ${\cal B}$,
where 
the coefficients $\alpha_i$ $(1,3,\ldots,6)$ and $\beta_j$ $(j=1,\ldots,4)$ 
as well as the coordinates are promoted to sections of some 
appropriate line bundles over the $\PP^1$ fibration. More precisely, 
we regard this $dP_9$ as a part of an elliptic $K3$ in the stable degeneration limit, 
which itself is fibered over ${\cal B}$ in such a way that the total space is an 
elliptic Calabi-Yau ${\cal Y}$, whose base ${\cal W}$ itself is a $\PP^1$ fibration 
over ${\cal B}$. 
We take 
\beqa
X&\in& \Gamma(({\cal L}\otimes {\cal N})^2),\n
Y&\in& \Gamma(({\cal L}\otimes {\cal N})^3),\n
v&\in& \Gamma({\cal N}),\n
u&\in& \Gamma({\cal L}^6),
\eeqa
where $\Gamma$ denotes the space of the sections, 
${\cal L}$ is the anti-canonical line bundle of the base ${\cal B}$, 
and ${\cal N}$ is the ``twisting'' line bundle over ${\cal B}$\footnote{
The first Chern class of ${\cal N}$ is customarily referred to as ``the $\eta$ class.''
} 
characterizing the vector bundle of the dual heterotic string theory 
compactified 
on an elliptic Calabi-Yau ${\cal Z}$ of complex 
dimension one less,  
whose complex structure is identical to that of the 
elliptic fibration at $z=\infty$. At the same time, the fiber of this elliptic 
fibration at infinity also plays the role of the ``dual'' torus, at which the 
values of rational sections of (\ref{E8W1123}) describe the spectral surface 
\cite{FMW}, {\it i.e.} 
the holonomies of the flat connections, and hence the moduli space of the 
heterotic vector bundle $V$\footnote{We assume $c_1(V)=0$ throughout this paper.}.
%
Then the affine coordinates $(z,x,y)$ transform as 
\beqa
z&\in& \Gamma({\cal M}),\n
x&\in& \Gamma(({\cal L}\otimes {\cal M})^2),\n
y&\in& \Gamma(({\cal L}\otimes {\cal M})^3)
\label{zxybundles}
\eeqa
as sections of line bundles over $\cal B$, 
where ${\cal M}\equiv{\cal L}^{-6}\otimes{\cal N}$. They are also 
sections of some line bundles over the $\PP^1$ with $(u:v)$ being its coordinates. 
Due to the Calabi-Yau condition for ${\cal Y}$, ${\cal W}$ is required 
to be such that the base ${\cal B}$ part of the anti-canonical class of ${\cal W}$ 
coincides with ${\cal L}\otimes{\cal M}$ (\ref{zxybundles}).

For example, if we take ${\cal B}$ to be $\PP^1$ and ${\cal Z}$ to be an 
elliptic $K3$, then ${}\cal L$ is an ${\cal O}(2)$ bundle whose sections are 
described by quadratic polynomials of the affine coordinate $z'$ of the 
base $\PP^1$. Also ${\cal N}$ is chosen to be an 
${\cal O}(12+n)={\cal L}^6\otimes {\cal L}^{\frac n2}$ bundle, 
corresponding to $12+n$ instantons 
for one of the two $E_8$ gauge groups of the six-dimensional heterotic 
string theory. 
${\cal M}$ is $ {\cal L}^{\frac n2}$. 
The corresponding dual F-theory description chooses 
\cite{MorrisonVafa} ${\cal W}$ to be a Hirzebruch surface $\FF_n$ 
so that the base $\PP^1$ part of the anti-canonical class is 
${\cal L}\otimes{\cal M}={\cal O}(2+n)$  
(or ${\cal O}(2-n)$ depending on the choice of the divisor ``at infinity''), 
geometrically realizing the twisting 
of the spectral cover of the heterotic dual. Then we are led to the well-known 
Weierstrass equation on a Hirzebruch surface \cite{MorrisonVafa}
 \beqa
0&=&y^2+x^3+f(z,z')x+g(z,z'),
\label{WeierstrassMV}\\
f(z,z')&=&f_{8+4n} + z f_{8+3n} + z^2 f_{8+2n} + z^3 f_{8+n} +\cdots, \n
g(z,z')&=&g_{12+6n} + z g_{12+5n} + z^2 g_{12+4n} + z^3 g_{12+3n} +\cdots,
\label{fgMV}
\eeqa
where $f_{8+(4-i)n}$ and $g_{12+(6-j)n}$ ($i,j=0,1,2,\ldots$) are polynomials 
of $z'$ with subscripts being their degrees in $z'$.
\footnote{In the {\em four}-dimensional compactifications of F-theory, 
one also needs to specify the so-called $\gamma$ class ($G$-flux), but 
it is irrelevant for the discussion here.}

%

In the general case, we write  (\ref{E8W1123}) 
in the Neron-Tate's form:
\beqa
0&=&y^2+x^3+(a_{1,0}+a_{1,1}z+a_{1,2}z^2+\cdots)xy\n
&&+(a_{2,0}+a_{2,1}z+a_{2,2}z^2+\cdots)x^2\n
&&+(a_{3,0}+a_{3,1}z+a_{3,2}z^2+\cdots)y\n
&&+(a_{4,0}+a_{4,1}z+a_{4,2}z^2+\cdots)x\n
&&+a_{6,0}+a_{6,1}z+a_{6,2}z^2+\cdots. 
\eeqa
The coefficients must be 
\beqa
f_{8+4n}=\beta_4=a_{4,0}
&&\in \Gamma({\cal L}^{-20}\otimes{\cal N}^4)~
=\Gamma({\cal L}^4\otimes {\cal M}^4)
\n
f_{8+3n}=\beta_3=a_{4,1}
&&\in \Gamma({\cal L}^{-14}\otimes{\cal N}^3)~
=\Gamma({\cal L}^4\otimes {\cal M}^3)
\n
f_{8+2n}=\beta_2=a_{4,2}
&&\in \Gamma({\cal L}^{-8}\otimes{\cal N}^2)~~
=\Gamma({\cal L}^4\otimes {\cal M}^2)
\n
f_{8+n}=\beta_1=a_{4,3}
&&\in \Gamma({\cal L}^{-2}\otimes{\cal N}^1)~~
=\Gamma({\cal L}^4\otimes {\cal M}^1)
\n
\n
g_{12+6n}=\alpha_6=a_{6,0}
&&\in \Gamma({\cal L}^{-30}\otimes{\cal N}^6)~
=\Gamma({\cal L}^6\otimes {\cal M}^6)
\n
g_{12+5n}=\alpha_5=a_{6,1}
&&\in \Gamma({\cal L}^{-24}\otimes{\cal N}^5)~
=\Gamma({\cal L}^6\otimes {\cal M}^5)
\n
g_{12+4n}=\alpha_4=a_{6,2}
&&\in \Gamma({\cal L}^{-18}\otimes{\cal N}^4)~
=\Gamma({\cal L}^6\otimes {\cal M}^4)
\n
g_{12+3n}=\alpha_3=a_{6,3}
&&\in \Gamma({\cal L}^{-12}\otimes{\cal N}^3)~
=\Gamma({\cal L}^6\otimes {\cal M}^3)
\n
(g_{12+2n}=\alpha_2=a_{6,4}
&&\in \Gamma({\cal L}^{-6}\otimes{\cal N}^2)~~
=\Gamma({\cal L}^6\otimes {\cal M}^2))
\n
g_{12+n}=\alpha_1=a_{6,5}
&&\in \Gamma({\cal L}^{-0}\otimes{\cal N}^1)~~
=\Gamma({\cal L}^6\otimes {\cal M}^1).
\label{dictionary}
\eeqa
In the above equations we have also displayed on the leftmost side  
the corresponding coefficient polynomials of the Weierstrass form for
the well-studied six-dimensional 
compactification. 

Now Looijenga's theorem states that the moduli space of the vector bundle is 
parameterized by the sections
\beqa
a_k&=&\Gamma({\cal L}^{-d_k}\otimes {\cal N}^{s_k})~~~(k=0,\ldots,{\rm rank}G),
\label{Looijenga'stheorem}
\eeqa
where $d_k$ is $0$ ($k=0$) or the degree of the independent Casimir of $G$, and 
$s_k$ is $1$ ($k=0$) or the coefficient of the $k$-th coroot when the lowest root $-\theta$ 
is expanded. 

In the present case, the minus of the powers of ${\cal L}$ 
in the middle column of (\ref{dictionary}) 
read
\beqa
0,2,8,12,14,18,20,24,30,
\eeqa
which precisely coincide with (``$0$'' and)
the set of degrees of independent Casimirs of $E_8$, 
while the powers of ${\cal N}$ are {\em close to} identical to 
the coefficients of the (co)root ($E_8$ is simply laced) expansion:
\beqa
-\theta&=&2\alpha_1+3\alpha_2+4\alpha_3+6\alpha_4+5\alpha_5+4\alpha_6
+3\alpha_7+2\alpha_8,
\eeqa
except that the power for $g_{12+n}=\alpha_1=a_{6,5}$, which should be 2, is 1. 
If $g_{12+2n}=\alpha_2=a_{6,4}$ were taken instead, then the power would 
become 2 which is correct, but then the power of ${\cal L}^{-1}$ would 
be 6 which does not agree with Looijenga's statement. 
Thus in this $E_8$ case, we have obtained 
 the weighted projective space 
${\rm W}\PP^8_{(1,2,2,3,3,4,4,5,6)}$
but (\ref{Looijenga'stheorem}) is not completely true \cite{FMW}.

\subsection{$E_7$ bundles by blowing up ($A_1$ singularity)}
In \cite{FMW} it was shown that $E_7$ bundles can be constructed 
in terms of a degree-4 equation in ${\rm W}\PP^3_{(1,1,1,2)}$. 
Sections of  Looijenga's weighted projective bundle are similarly 
parameterized by the coefficients of the Weierstrass equation, 
which themselves are sections of a particular set of line bundles 
specified by  Looijenga's theorem (\ref{Looijenga'stheorem}).
In this section we will show that these setups naturally arise by 
blowing up the singularity on the degree-6 equation in 
${\rm W}\PP^3_{(1,1,2,3)}$ for $E_8$ discussed in the previous 
section, according to a well-known procedure known as 
Tate's algorithm.

Physically, an $E_7$ bundle implies an $SU(2)$ unbroken gauge symmetry 
of one of the two $E_8$ of heterotic string theory. Mathematically, 
this is a reflection of the structure of the Mordell-Weil lattice \cite{OguisoShioda} 
stating the complementarity in $E_8$ 
of the sections and the singularities of a rational elliptic surface $dP_9$.

Below we use in the process of blowing up, 
{\em even in the general case not restricted to the six-dimensional case}, 
the notation:
\beqa
f_{8+(4-i)n}&:=&a_{4,i}, \\
g_{12+(6-j)n}&:=&a_{6,j},  
\eeqa
by using the dictionary (\ref{dictionary}).
\footnote{\label{dands}More generally, a degree $(an+b)$ polynomial 
in $z'$ in the 6D F-theory compactification corresponds to 
a section of ${\cal L}^{-d}\otimes {\cal N}^{s}$ with 
$d=6a-\frac b2$ and $s=a$.} 
In that case, $n$ no longer has 
the meaning of the number of instantons, but is rather just 
a dummy variable with its coefficient 
specifying the powers of the twisting line bundle to which 
the section belong. 
This notation is intended for the convenience of, and will be particularly useful 
to,  the readers who are 
familiar with the well-known six-dimensional F-theory compactification 
\cite{MorrisonVafa,BIKMSV}.  This enables us to easily recognize that the 
sections parameterizing a Looijenga's weighted projective space 
are nothing but the set of independent polynomials parameterizing 
the complex structure of the elliptic manifold in F-theory with a singularity, 
which is the orthogonal complement in $E_8$ of the gauge group of 
the vector bundle. Why they are the orthogonal complement of each 
other will be explained in section 4.

In order to have an $SU(2)=A_1=I_2$ singularity, we assume that 
the coefficients $f_{8+4n}$, $g_{12+6n}$ and $g_{12+5n}$ in (\ref{fgMV}) 
can be written 
in term of some $h_{2n+4}\in\Gamma({\cal L}^2\otimes{\cal M}^2)$ as 
\beqa
f_{8+4n}&=&-3h_{2n+4}^2,\n
g_{12+6n}&=&2h_{2n+4}^3,\n
g_{12+5n}&=&-h_{2n+4} f_{8+3n}.
\eeqa
Then the discriminant 
\beqa
\Delta&=&4 f(z,z')^3 + 27 g(z,z')^2
\eeqa
becomes $O(z^2)$ or higher at $z=0$, implying an $A_1$ 
singularity. The location of the singularity is $y=z=0$ but $x=h_{2n+4}$, 
so it is not at the origin in general. So we define 
\beqa
x_{new}&\equiv&x-h_{2n+4},
\eeqa
then (\ref{WeierstrassMV}) becomes
\beqa
0&=&y^2+x_{new}^3+3 h_{2n+4}x_{new}^2 
+(f_{8+3n}z+f_{8+2n}z^2+f_{8+n}z^3+f_{8}z^4)x_{new}\n
&&
+(h_{2n+4}f_{8+2n}+g_{12+4n})z^2
+(h_{2n+4}f_{8+n}+g_{12+3n})z^3\n
&&
+(h_{2n+4}f_{8}+g_{12+2n})z^4
+g_{12+n}z^5
+g_{12}z^6.
\label{shiftedWeierstrass}
\eeqa
By construction it has a singularity at $x_{new}=y=z=0$ so we 
blow it up at 
$(x_{new},y,z)=(0,0,0)\in \CC^3$ by defining  
\beqa
\tilde{\CC}^3&=&
\left\{
\left((x_{new},y,z),({\xi:\eta:\zeta})\right)\in\CC^3\times\PP^2 | (x_{new},y,z)\in({\xi:\eta:\zeta})
\right\},
\label{blowupofC3}
\eeqa
where $ (x_{new},y,z)\in({\xi:\eta:\zeta})$ means that $(x_{new},y,z)$ and 
$({\xi:\eta:\zeta})$ are parallel to each other. 

Let $x'\equiv\frac\xi\zeta$, $y'\equiv\frac\eta\zeta$ 
in the affine patch with $\zeta\neq 0$, then 
\beqa
(x_{new},y,z)&=&(x'z,y'z,z).
\label{xprimeyprime}
\eeqa
Plugging this into (\ref{shiftedWeierstrass}) and dividing it by $z^2$, 
we have
\beqa
0&=& {y'}^2+{x'}^3 z+3 h_{2n+4}{x'}^2 
+(f_{8+3n}+f_{8+2n}z+f_{8+n}z^2+f_{8}z^3) x'\n
&&
+(h_{2n+4}f_{8+2n}+g_{12+4n})
+(h_{2n+4}f_{8+n}+g_{12+3n})z\n
&&
+(h_{2n+4}f_{8}+g_{12+2n})z^2
+g_{12+n}z^3
+g_{12}z^4.
\label{blownupWeierstrass}
\eeqa
One can show that this is a smooth curve {\em unless} 
\beqa
f_{8+3n}^2 -12 h_{2n+4}(h_{2n+4}f_{8+2n}+g_{12+4n})&=&0
\label{xiDiscriminant}
\eeqa
is satisfied. 
In fact, the left hand side of (\ref{xiDiscriminant}) is the coefficient of the $O(z^2)$ 
term of the discriminant $\Delta$, so that (\ref{xiDiscriminant}) implies 
$ord(\Delta)\geq 3$ at $z=0$.
Note that this is a necessary condition for the curve to be singular, and even if 
(\ref{xiDiscriminant}) holds, (\ref{blownupWeierstrass}) still remains regular 
unless some additional conditions 
are satisfied, as we see in the next section.

Now we observe that 
(\ref{blownupWeierstrass}) is nothing but a degree-4 equation in 
${\rm W}\PP^3_{(1,1,1,2)}$ 
($(u,v,X',Y')\sim(\lambda u,\lambda v,\lambda X',\lambda^2 Y')$):
\beqa
0&=&Y'^2+X'^3v+a_{2,0}X'^2 u^2\n
&&+(a_{4,1} u^3+a_{4,2} u^2 v+ a_{4,3} u v^2 +a_{4,4}  v^3)X'\n
&&+(a_{6,2} u^4 +a_{6,3} u^3 v +a_{6,4} u^2 v^2 +a_{6,5} u v^3 +a_{6,6} v^4)
\label{E7W1112}
\eeqa
expressed in the affine patch with $u\neq 0$ in terms of the affine coordinates 
$(u,v,X',Y')\sim (1,\frac v u,\frac {X'}{u},\frac {Y'}{u^2})
\equiv(1,z,x',y')$ :
\beqa
0&=&y'^2+x'^3z+a_{2,0}x'^2 \n
&&+(a_{4,1} +a_{4,2}  z+ a_{4,3}  z^2 +a_{4,4}  z^3)x'\n
&&+(a_{6,2}  +a_{6,3}  z +a_{6,4}  z^2 +a_{6,5}  z^3 +a_{6,6} z^4)
\label{E7W1112MV}
\eeqa
with 
\beqa
a_{2,0}&=&3 h_{2n+4}\n
a_{4,1}&=&f_{8+3n}\n
a_{4,2}&=&f_{8+2n}\n
a_{4,3}&=&f_{8+n}\n
a_{4,4}&=&f_{8}\n
a_{6,2}&=&h_{2n+4}f_{8+2n}+g_{12+4n}\n
a_{6,3}&=&h_{2n+4}f_{8+n}+g_{12+3n}\n
a_{6,4}&=&h_{2n+4}f_{8}+g_{12+2n}\n
a_{6,5}&=&g_{12+n}\n
a_{6,6}&=&g_{12}.
\eeqa

The degree-4 equation in 
${\rm W}\PP^3_{(1,1,1,2)}$ 
(\ref{E7W1112}) is precisely the one found in \cite{FMW} for $E_7$ bundles. 
Thus we see that the set up for the construction of $E_7$ bundles in \cite{FMW} 
is naturally obtained by blowing up the singularity of the Weierstrass equation 
in ${\rm W}\PP^3_{(1,1,2,3)}$ for $E_8$ bundles. 

Since
\beqa
a_{i,j}&\in&{\cal L}^i\otimes {\cal M}^{i-j}~=~{\cal L}^{6j-5i}\otimes({\cal L}^6\otimes {\cal M})^{i-j},
\eeqa
we have
\beqa
a_{2,0}&\in&\Gamma({\cal L}^{-10}\otimes({\cal L}^6\otimes {\cal M})^{2})\n
a_{4,1}&\in&\Gamma({\cal L}^{-14}\otimes({\cal L}^6\otimes {\cal M})^{3})\n
a_{4,2}&\in&\Gamma({\cal L}^{-8}\otimes({\cal L}^6\otimes {\cal M})^{2})\n
a_{4,3}&\in&\Gamma({\cal L}^{-2}\otimes({\cal L}^6\otimes {\cal M})^{1})\n
a_{6,2}&\in&\Gamma({\cal L}^{-18}\otimes({\cal L}^6\otimes {\cal M})^{4})\n
a_{6,3}&\in&\Gamma({\cal L}^{-12}\otimes({\cal L}^6\otimes {\cal M})^{3})\n
a_{6,4}&\in&\Gamma({\cal L}^{-6}\otimes({\cal L}^6\otimes {\cal M})^{2})\n
a_{6,5}&\in&\Gamma({\cal L}^{-0}\otimes({\cal L}^6\otimes {\cal M})^{1}).
\label{E7sections}
\eeqa
Note that $a_{4,4}\in\Gamma({\cal L}^4)$ or $a_{6,6}\in\Gamma({\cal L}^6)$ 
does not have information of the vector bundles but describes the 
complex structure of the Calabi-Yau on the heterotic side, and that of 
the elliptic fibration connecting the two $dP_9$ fibrations on the F-theory side.

We see in (\ref{E7sections}) that the minus of the powers of ${\cal L}$ 
read 
 \beqa
0,2,6,8,10,12,14,18,
\eeqa
which coincides with the set of degrees of independent Casimirs of $E_7$ (including $0$),
and the powers of ${\cal N}$ are the coefficients of the expansion of the highest root of $E_7$:
\beqa
-\theta&=&2\alpha_1+2\alpha_2+3\alpha_3+4\alpha_4+3\alpha_5
+2\alpha_6+1\alpha_7,
\eeqa
being (this time) in complete agreement with Looijenga.


\subsection{$E_6$ bundles ($A_2$ singularity)}
Next we suppose that (\ref{xiDiscriminant}) is satisfied.
Since the first term is a square of a section, so must be the second term. 
This is achieved by requiring 
\beqa
h_{2n+4}&=&h_{n+2}^2\n
f_{8+3n}&=&12 h_{n+2}H_{2n+6}\n
g_{12+4n}&=&12H_{2n+6}^2-h_{n+2}^2 f_{8+2n}
\label{E6factorization}
\eeqa
for some $h_{n+2}$ and $H_{2n+6}$.
These conditions are the ones for the exceptional curve to factorize into 
two lines, and the singularity becomes $I_3$ of the Kodaira classification. 
This is smooth unless 
\beqa
-2 H_{2n+6}( h_{n+2}^2 f_{8+2n} +4 H_{2n+6}^2)
+h_{n+2}^3(h_{n+2}^2 f_{8+n}+g_{12+3n})&=&0
\label{D5singularcondition}
\eeqa
is satisfied, in which the order of the discriminant would become 
higher than 3 
and we would need a further blow-up. 
Plugging 
(\ref{E6factorization}) 
into (\ref{shiftedWeierstrass}), we find 
\beqa
0&=&y^2+x_{new}^3+3 h_{n+2}^2x_{new}^2 
+(12 h_{n+2}H_{2n+6}z+f_{8+2n}z^2+f_{8+n}z^3+f_{8}z^4)x_{new}\n
&&
+12H_{2n+6}^2 z^2
+(h_{n+2}^2f_{8+n}+g_{12+3n})z^3
+(h_{n+2}^2f_{8}+g_{12+2n})z^4
+g_{12+n}z^5
+g_{12}z^6.\n
\eeqa
We can further rewrite it in terms of
\beqa
y_{new}&\equiv&y-\sqrt{3} i (h_{n+2} x_{new}+2 H_{2n+6}z)
\eeqa
as 
\beqa
0&=&y_{new}^2+x_{new}^3
+2\sqrt{3} i h_{n+2} x_{new} y_{new} + 4\sqrt{3} i H_{2n+6} z y_{new}
+(f_{8+2n}z^2+f_{8+n}z^3+f_{8}z^4)x_{new}\n
&&
+(h_{n+2}^2f_{8+n}+g_{12+3n})z^3
+(h_{n+2}^2f_{8}+g_{12+2n})z^4
+g_{12+n}z^5
+g_{12}z^6.
\eeqa
Note that $x_{new}=y=0 \Leftrightarrow x_{new}=y_{new}=0$ at $z=0$.
Similarly to (\ref{xprimeyprime}) in (\ref{shiftedWeierstrass}), we set 
\beqa
(x_{new},y_{new},z)&=&(x'z,y'z,z)
\label{xprimeyprime2}
\eeqa
to find 
\beqa
0&=& {y'}^2+{x'}^3 z+2\sqrt{3} i h_{n+2}x' y' + 4\sqrt{3} i H_{2n+6}y'
+(f_{8+2n}z+f_{8+n}z^2+f_{8}z^3) x'\n
&&
+(h_{n+2}^2f_{8+n}+g_{12+3n})z
+(h_{n+2}^2f_{8}+g_{12+2n})z^2
+g_{12+n}z^3
+g_{12}z^4.
\label{blownupWeierstrass2}
\eeqa
Again, this is a fourth-order equation in 
${\rm W}\PP^3_{(1,1,1,2)}$ 
($(u,v,X',Y')\sim(\lambda u,\lambda v,\lambda X',\lambda^2 Y')$):
\beqa
0&=&Y'^2+X'^3v+a_{1,0}X'Y'u  + a_{3,1}Y' u^2\n
&&+(a_{4,2} u^2 v+ a_{4,3} u v^2 +a_{4,4}  v^3)X'\n
&&+a_{6,3} u^3 v +a_{6,4} u^2 v^2 +a_{6,5} u v^3 +a_{6,6} v^4
\label{E6W1112}
\eeqa
expressed in terms of the affine coordinates 
$(u,v,X',Y')\sim (1,\frac v u,\frac {X'}{u},\frac {Y'}{u^2})
\equiv(1,z,x',y')$ in the patch
$u\neq 0$:
\beqa
0&=&y'^2+x'^3z+a_{1,0}x'y'  + a_{3,1}y' \n
&&+(a_{4,2}  z+ a_{4,3}  z^2 +a_{4,4}  z^3)X'\n
&&+a_{6,3}  z +a_{6,4}  z^2 +a_{6,5}  z^3 +a_{6,6} z^4
\label{E6W1112MV}
\eeqa
with
\beqa
a_{1,0}&=&2\sqrt{3} i h_{n+2}\n
a_{3,1}&=&4\sqrt{3} i H_{2n+6}\n
a_{4,2}&=&f_{8+2n}\n
a_{4,3}&=&f_{8+n}\n
a_{4,4}&=&f_{8}\n
a_{6,3}&=&h_{2n+4}f_{8+n}+g_{12+3n}\n
a_{6,4}&=&h_{2n+4}f_{8}+g_{12+2n}\n
a_{6,5}&=&g_{12+n}\n
a_{6,6}&=&g_{12}.
\eeqa

In this case we have
\beqa
a_{1,0}&\in&\Gamma({\cal L}^{-5}\otimes({\cal L}^6\otimes {\cal M})^{1})\n
a_{3,1}&\in&\Gamma({\cal L}^{-9}\otimes({\cal L}^6\otimes {\cal M})^{2})\n
a_{4,2}&\in&\Gamma({\cal L}^{-8}\otimes({\cal L}^6\otimes {\cal M})^{2})\n
a_{4,3}&\in&\Gamma({\cal L}^{-2}\otimes({\cal L}^6\otimes {\cal M})^{1})\n
a_{6,3}&\in&\Gamma({\cal L}^{-12}\otimes({\cal L}^6\otimes {\cal M})^{3})\n
a_{6,4}&\in&\Gamma({\cal L}^{-6}\otimes({\cal L}^6\otimes {\cal M})^{2})\n
a_{6,5}&\in&\Gamma({\cal L}^{-0}\otimes({\cal L}^6\otimes {\cal M})^{1}),
\eeqa
which is consistent with the facts that the degrees of the independent 
Casimirs of $E_6$ (including 0) are 
\beqa
0,2,5,6,8,9,12
\eeqa
and the (co)root expansion of the highest root is
\beqa
-\theta&=&1\alpha_1+2\alpha_2+2\alpha_3+3\alpha_4+2\alpha_5
+1\alpha_6.
\eeqa

Thus we have seen that not only $E_7$ bundles but 
$E_6$ bundles can also be constructed from 
${\rm W}\PP^3_{(1,1,1,2)}$. 
In contrast, instead of ${\rm W}\PP^3_{(1,1,1,2)}$, 
${\rm W}\PP^3_{(1,1,1,1)}$ was used in FMW \cite{FMW}, 
which can be obtained by a further blow-up as we will see in the next section. 
Note, however, that in the case of the $I_3=A_2$ singularity 
the exceptional curve arising in the $I_2=A_1$ singularity simply 
splits into to two lines, in which no additional blow-ups are needed, 
and therefore ${\rm W}\PP^3_{(1,1,1,2)}$ suffice. Of course, 
one is free to blow it up so it is not a contradiction.

\subsection{$D_5$ bundles ($A_3$ singularity)}
In this section we consider the case in which
(\ref{D5singularcondition}) is satisfied and the curve in the previous section 
becomes singular.  Then the
discriminant $\Delta$ bedomes $ord(\Delta)\geq 4$. 
In this case we require 
\beqa
H_{2n+6}&=&h_{n+2}H_{n+4}
\eeqa
for some $H_{n+4}$. Due to 
(\ref{D5singularcondition}),  we need to have 
\beqa
g_{12+3n}&=& 2H_{n+4}(f_{8+2n}+4H_{n+4}^2)-h_{n+2}^2 f_{8+n}.
\eeqa
Then (\ref{blownupWeierstrass2}) becomes singular at 
$x'=-2H_{n+4}$, $y'=z=0$. To resolve this singularity we define 
\beqa
x'_{new}&\equiv&x'+2H_{n+4},
\eeqa
then 
\beqa
0&=&{y'}^2+{x'_{new}}^3 z+2 i \sqrt{3} h_{n+2}x'_{new}y'-6 z H_{n+4}{x'_{new}}^2\n
&&+x'_{new} \left(z \left(f_{8+2 n}+12 H_{n+4}^2\right)+f_{8+n}z^2 +f_8 z^3\right)\n
&&+z^2 \left(
   h_{n+2}^2 f_8 -2 H_{n+4}f_{8+n} +g_{12+2n}\right)\n
&&   +z^3 \left(-2 H_{n+4}f_8 +g_{12+n}\right)+ g_{12}z^4.
\label{blownupWeierstrassD5}
\eeqa
The singularity is located at $x'_{new}=y'=z=0$, so defining 
\beqa
y'&=&\tilde{y}'z
\label{y'scaling}
\eeqa
and factoring $z$ out, we derive 
\beqa
0&=&\tilde{y}'{}^2 z
+{x'_{new}}^3 +f_8 x'_{new} z^2+ g_{12}z^3\n
&&+2 i \sqrt{3} h_{n+2}x'_{new}{\tilde{y}}'
-6 H_{n+4}{x'_{new}}^2\n
&&+x'_{new} \left(\left(f_{8+2 n}+12 H_{n+4}^2\right)+f_{8+n}z \right)\n
&&+z \left(
   h_{n+2}^2 f_8 -2 H_{n+4}f_{8+n} +g_{12+2n}\right)\n
&&   +z^2 \left(-2 H_{n+4}f_8 +g_{12+n}\right).
\eeqa
Rewriting this equation as 
\beqa
0&=&\tilde{y}'{}^2 z
+{x'_{new}}^3 +a_{4,4} x'_{new} z^2+ a_{6,6}z^3\n
&&+a_{1,0}x'_{new}{\tilde{y}}'
+a_{2,1}{x'_{new}}^2\n
&&+x'_{new} \left(a_{4,2}+a_{4,3}z \right)\n
&&+a_{6,4} z   +a_{6,5}z^2,  
\eeqa
we see that this is a third-order equation in 
${\rm W}\PP^3_{(1,1,1,1)}$ 
($(u,v,X',\tilde Y')\sim(\lambda u,\lambda v,\lambda X',\lambda \tilde Y')$):
\beqa
0&=&\tilde{Y}'{}^2 v
+{X'}^3 +a_{4,4} X' v^2+ a_{6,6}v^3\n
&&+u^2 \left(a_{4,2}X' +a_{6,4} v\right)\n
&&+u\left(a_{1,0}X'{\tilde{Y}}'
+a_{2,1}{X'}^2 +a_{4,3}X' v+a_{6,5} v^2 \right),
\label{D5W1111}
\eeqa
expressed in terms of the affine coordinates 
$(u,v,X',\tilde Y')\sim (1,\frac v u,\frac {X'}{u},\frac {\tilde Y'}{u})
\equiv(1,z,x'_{new},\tilde y')$
in the patch $u\neq 0$, with the identifications 
\beqa
a_{1,0}&=&2\sqrt{3} i h_{n+2}\n
a_{2,1}&=&-6 H_{n+4}\n
a_{4,2}&=&f_{8+2n}+12 H_{n+4}^2\n
a_{4,3}&=&f_{8+n}\n
a_{6,4}&=&h_{n+2}^2 f_8 -2 H_{n+4}f_{8+n} +g_{12+2n}\n
a_{6,5}&=&-2 H_{n+4}f_8 +g_{12+n}
\label{D5a's}
\eeqa
($a_{4,4}=f_{8}$, $a_{6,6}=g_{12}$).

The relevant sections are 
\beqa
a_{1,0}&\in&\Gamma({\cal L}^{-5}\otimes({\cal L}^6\otimes {\cal M})^{1})\n
a_{2,1}&\in&\Gamma({\cal L}^{-4}\otimes({\cal L}^6\otimes {\cal M})^{1})\n
a_{4,2}&\in&\Gamma({\cal L}^{-8}\otimes({\cal L}^6\otimes {\cal M})^{2})\n
a_{4,3}&\in&\Gamma({\cal L}^{-2}\otimes({\cal L}^6\otimes {\cal M})^{1})\n
a_{6,4}&\in&\Gamma({\cal L}^{-6}\otimes({\cal L}^6\otimes {\cal M})^{2})\n
a_{6,5}&\in&\Gamma({\cal L}^{-0}\otimes({\cal L}^6\otimes {\cal M})^{1}),
\eeqa
which agree with the degrees of Casimirs of $D_5$ with 0:
\beqa
0,2,4,5,6,8
\eeqa
and the coroot expansion:
\beqa
-\theta&=&1\alpha_1+2\alpha_2+2\alpha_3+1\alpha_4+1\alpha_5.
\eeqa
Thus we have derived  ${\rm W}\PP^5_{(1,1,1,1,2,2)}$ from 
a third-order equation in 
${\rm W}\PP^3_{(1,1,1,1)}$. This construction of $D_5$ bundles was 
not explicitly mentioned in FMW.

\subsection{$A_4$ bundles ($A_4$ singularity)}
If we further assume 
\beqa
f_{8+2n}&=&-12H_{n+4}^2+12 h_{n+2}p_{n+6},\n
g_{12+2n}&=&12 p_{n+6}^2+2 f_{8+n}H_{n+4}-f_8 h_{n+2}^2
\eeqa
for some $p_{n+6}$ in (\ref{blownupWeierstrassD5}), we have
$ord(\Delta)\geq 5$ and the exceptional curve again splits 
into two lines. 
In this case, unlike the case for $E_6$ bundles, 
the singularity of (\ref{blownupWeierstrassD5}) is not resolved by 
(\ref{y'scaling}) but we also need to scale $x'_{new}$. 
This can be done, but we can still use (\ref{D5W1111}) to 
see which sections are independent. 
Then 
(\ref{D5a's}) reads
\beqa
a_{1,0}&=&2\sqrt{3} i h_{n+2}\n
a_{2,1}&=&-6 H_{n+4}\n
a_{4,2}&=&12 h_{n+2} p_{n+6}\n
a_{4,3}&=&f_{8+n}\n
a_{6,4}&=&12 p_{n+6}^2\n
a_{6,5}&=&-2 H_{n+4}f_8 +g_{12+n},
\label{A4a's}
\eeqa
where we see that $a_{4,2}$ and $a_{6,4}$ are simplified. 
They are the coefficients of  the $u^2$ term in (\ref{D5W1111}) so 
using 
\beqa
a_{3,2}= 4\sqrt{3}i p_{n+6}, ~~~\tilde{Y}'_{new}=\tilde{Y}'-2\sqrt{3}iu p_{n+6}
\eeqa
we have
\beqa
0&=&\tilde{Y}'{}_{new}^2 v +a_{3,2}\tilde{Y}'{}_{new} uv
+{X'}^3 +a_{4,4} X' v^2+ a_{6,6}v^3\n
&&+u\left(a_{1,0}X'{\tilde{Y}}'
+a_{2,1}{X'}^2 +a_{4,3}X' v+a_{6,5} v^2 \right).
\label{A4W1111}
\eeqa
$a_{4,2}$ and $a_{6,4}$ in (\ref{D5W1111}) are thus eliminated. 
In this way, for $A_4$ bundles, we have obtained a third-order equation in
${\rm W}\PP^3_{(1,1,1,1)}$ (which is singular but can be smooth by a 
blow up) with
\beqa
a_{1,0}&\in&\Gamma({\cal L}^{-5}\otimes({\cal L}^6\otimes {\cal M})^{1})\n
a_{2,1}&\in&\Gamma({\cal L}^{-4}\otimes({\cal L}^6\otimes {\cal M})^{1})\n
a_{3,2}&\in&\Gamma({\cal L}^{-3}\otimes({\cal L}^6\otimes {\cal M})^{1})\n
a_{4,3}&\in&\Gamma({\cal L}^{-2}\otimes({\cal L}^6\otimes {\cal M})^{1})\n
a_{6,5}&\in&\Gamma({\cal L}^{-0}\otimes({\cal L}^6\otimes {\cal M})^{1}).
\eeqa
Again this agrees with the set of Casimirs of $A_4$ with degrees (with 0): 
\beqa
0,2,3,4,5
\eeqa
and the expansion
\beqa
-\theta&=&1\alpha_1+1\alpha_2+1\alpha_3+1\alpha_4.
\eeqa

\subsection{$A_3,A_2,A_1$ bundles ($D_5,E_6,E_7$ singularity)}
 
So far we have considered bundles for the $E$ series up to $E_4=A_4$. 
Since $E_3$ or $E_2$ is not a simple Lie algebra, we need a separate discussion 
for them. Instead, however, $A_3$, $A_2$ and $A_1$ bundles can 
be similarly constructed by setting $h_{n+2}$, $H_{n+4}$ and 
$p_{n+6}$ to zero in this order. In either case, one can show that there is an 
agreement between the powers of the line bundles and the degrees of 
the independent Casimirs and the expansion coefficients of the 
highest weight.  Note that in these cases there is still a 
singularity at $z=0$ to be further blown up.

\section{Relation to the independent polynomials characterizing the complex structure}

In the preceding sections we have seen that 
for $E_7$, $E_6$, $D_5$, $A_4$, $A_3$, $A_2$ and $A_1$ bundles 
(besides $E_8$ bundles which are exceptional) 
the necessary sections which constitute the corresponding weighted 
projective space stated in Looijenga's theorem are naturally 
obtained by a series of singularity enhancements of the elliptic manifold 
followed by the blowing-up procedure. We can now notice that they are
nothing but 
the four-dimensional analogue 
of the set of independent polynomials in the six-dimensional F-theory 
\cite{MorrisonVafa,BIKMSV}
parameterizing the complex structure of the elliptic manifold. 
The type of the singularity is always the one 
orthogonal 
to the gauge group of the vector bundle  in the whole $E_8$. 
Indeed, as shown in TABLE I, 
there is a perfect correspondence between the set of independent 
polynomials describing the complex structure in 6D and the set of 
numbers  
$d$ and $s$ characterizing the sections 
required by Looijenga's theorem, for all the cases of the bundle groups 
discussed in the preceding section, as well as the other cases 
for simple, simply-laced gauge groups listed in 
\cite{BIKMSV}. As we already noted in the previous footnote \ref{dands}, 
a degree $(an+b)$ polynomial 
in $z'$ 
corresponds to a section of ${\cal L}^{-d}\otimes {\cal N}^{s}$ with 
$d=6a-\frac b2$ and $s=a$. 

For $D_4$ bundles, which are not discussed in the previous section, 
we consider curves with a $D_4$ singularity. This can be obtained 
by restricting $h_{n+2}$ and $H_{2n+6}$ to be zero 
in the $A_2$ curve 
(used for $E_6$ bundles) and requiring   
the sixth-order term of the discriminant to be of the form \cite{BIKMSV,Tani}
\beqa
4 f_{2n+8}^3 + 27 g_{3n+12}^2&=&j_{n+4}^2 k_{n+4}^2 (j_{n+4}^2+k_{n+4}^2)
\eeqa
for some $j_{n+4}$ and $k_{n+4}$, which are precisely the polynomials 
with correct degrees needed to constitute the weighted projective space.

$D_6$ and $A_5$ bundles, 
which also do not appear in the previous section, are 
interesting because they are the cases where the singularity 
has two non-abelian factors. 
For $D_6$ bundles, 
one can show that the relevant curve is, 
again in the 6D notation, 
\beqa
0&=&y^2 + x^3 + 3(h_{2n+4}+h_{n+4}z)x^2 \n
&&+z(z+h_n)(p_{2n+8} + q_{n+8} z + s_8 z^2)x\n
&&+z^2(z+h_n)^2 (v_{2n+12} + w_{n+12} z + y_{12} z^2).
\label{D6bundlecurve}
\eeqa
This curve has an $A_1\times A_1$ ($SU(2)\times SU(2)$) singularity.
The lines $z=0$ and $z+h_n=0$ are the loci of the 7-branes 
responsible for the two unbroken $SU(2)$ gauge symmetries. 
Indeed, the discriminant takes the forms 
\beqa
\Delta&=&z^2 h_n^2 h_{2n+4}^2 K_{4n+16}+O(z^3)\n
&=&\tilde z^2h_n^2 \tilde h_{2n+4}^2 \tilde K_{4n+16}+O(\tilde z^3),
\label{D6bundlediscriminant}
\eeqa
where $\tilde z=z+h_n$ and $\tilde h_{2n+4}=h_{2n+4}-h_{n+4} h_n$. 
$K_{4n+16}$ and $\tilde K_{4n+16}$ are given by  
\beqa
K_{4n+16} &=& 9(12h_{2n+4}v_{2n+12}-p_{2n+8}^2),\n
\tilde K_{4n+16}&=& 9(12\tilde h_{2n+4}\tilde v_{2n+12}-\tilde p_{2n+8}^2),
\label{K4n16}
\eeqa
where $\tilde h_{2n+4}$, $\tilde v_{2n+12}$ and $\tilde p_{2n+8}$ are 
the coefficient polynomials appearing when (\ref{D6bundlecurve}) 
is re-expressed in terms of $\tilde z$ as
\beqa
0&=&y^2 + x^3 + 3(\tilde h_{2n+4}+h_{n+4}\tilde z)x^2 \n
&&+\tilde z(\tilde z-h_n)(\tilde p_{2n+8} + \tilde q_{n+8} \tilde z + s_8 \tilde z^2)x\n
&&+\tilde z^2(\tilde z-h_n)^2 (\tilde v_{2n+12} + \tilde w_{n+12} \tilde z + y_{12} \tilde z^2).
\eeqa
(\ref{D6bundlediscriminant}) is consistent with the fact that 
the 6D heterotic charged matter 
consists of $n$ $({\bf 2},{\bf 2})$ and $4n+16$ 
($({\bf 2},{\bf 1})\oplus({\bf 1},{\bf 2})$) computed by the index theorem.
Note that the loci of $h_{2n+4}$ and $\tilde h_{2n+4}$ do not contribute 
to charged matter since the enhanced fiber type there is 
$III$ in the Kodaira classification so the singularity type is unchanged.  
 One can also verify that, in the six-dimensional case, 
the total number of degrees of freedom of these polynomials   
\beqa
(n+13)+(n+9)+(n+5)+(n+1)+(2n+13)+(2n+9)+(2n+5)-1
\eeqa
is equal to $10n+54$ which precisely matches the number of 
neutral hypermultiplets. 
We can see that the sections $w_{n+12}$, $q_{n+8}$,
$h_{n+4}$, $h_n$, $v_{2n+12}$, $p_{2n+8}$ and $h_{2n+4}$
are precisely the polynomials expected to arise by Looijenga's 
theorem as are shown in TABLE I.

Similarly, the curve for an $A_5$ bundle is given by 
\beqa
0&=&y^2 + x^3 + 3(h_{n+2}^2+h_{n+4}z)x^2 \n
&&+z(z+h_n)(12 h_{n+2}v_{n+6} + q_{n+8} z + s_8 z^2)x\n
&&+z^2(z+h_n)^2 (12 v_{n+6}^2 + w_{n+12} z + y_{12} z^2),
\eeqa
which has an $E_3=A_2\times A_1$ ($SU(3)\times SU(2)$) singularity.
Here the $O(z^2)$ term in \eqref{D6bundlediscriminant} vanishes
($K_{4n+16}=0$ in \eqref{K4n16}) and the $A_1$ singularity at $z=0$ is enhanced to $A_2$. 
The discriminant in this case is
\beqa
\Delta&=&z^3 h_n^2 h_{n+2}^3 K_{4n+18}+O(z^4)\n
&=&\tilde z^2 h_n^3 \tilde h_{2n+4}^2 \tilde K_{3n+16}+O(\tilde z^3),
\eeqa
being in agreement with the fact that the 6D heterotic charged matter 
hypermultiplets are $\frac n2$ ($({\bf 3},{\bf 2})\oplus(\bar{\bf 3},{\bf 2}))$, $2n+9$ 
($({\bf 3},{\bf 1})\oplus(\bar{\bf 3},{\bf 1})$) and $3n+16$ $({\bf 1},{\bf 2})$.
The number of degrees of freedom of the polynomials also 
agrees with the number of neutral hypermultiplets $6n+37$.
Again, the sections $w_{n+12}$, $q_{n+8}$,
$v_{n+6}$, $h_{n+4}$, $h_{n+2}$ and $h_{n}$
have the desired set of $d$ and $s$ as are shown in TABLE I.

Finally, let us consider $E_3 = SU(3)\times SU(2)$ bundles and $SU(2)\times SU(2)$ bundles.
These groups are the orthogonal complements of $A_5 = SU(6)$ and $D_6 = SO(12)$ in $E_8$. 
Although these are not simple groups (and hence outside the assumption 
of Looijenga's theorem), it is interesting to examine 
whether or not a similar characterization of the bundles is possible 
in these cases.\footnote{$E_2$ contains $U(1)$ and hence is beyond 
the scope of this paper.}

For $E_3 = SU(3)\times SU(2)$ bundles, we consider curves with a $A_5$ singularity.
It is realized by further tuning the complex structure of the $A_4$ singularity ($A_4$ bundles)
parametrized by the polynomials (\ref{A4a's}) to the following special forms: 
\beqa
h_{n+2}&=&\tilde h_{n+2-r} t_r,\n
H_{n+4}&=&\tilde H_{n+4-r} t_r, \n
p_{n+6}&=&\tilde h_{n+2-r} u_{r+4},\n
f_{n+8}&=&\tilde f_{n+8-r} t_r -12 \tilde H_{n+4-r} u_{r+4},\n
g_{n+12}&=&2\tilde f_{n+8-r} u_{r+4} +2 f_8 H_{n+4}
\label{hHpfg}
\eeqa
for some $h_{n+2-r}$, $t_r$, 
$\tilde H_{n+4-r}$, 
$u_{r+4}$ and 
$\tilde f_{n+8-r}$, 
which describes the heterotic configuration with $4+r$ of 
$12+n$ instantons are in $SU(2)$ in $E_3$ 
and the remaining $8+n-r$ are in $SU(3)$.
Apparently, besides $f_8$ which describes the complex 
structure of the heterotic Calabi-Yau manifold, these five 
sections are needed to parametrize the moduli space of 
the bundle. However, defining 
\beqa
p_{n+6}&\equiv&\tilde h_{n+2-r} u_{r+4},\n
f^{(1)}_{n+8}&\equiv&\tilde f_{n+8-r} t_r,\n
f^{(2)}_{n+8}&\equiv&\tilde H_{n+4-r} u_{r+4},\n
g'_{n+12}&\equiv&2\tilde f_{n+8-r} u_{r+4},
\label{pf1f2g}
\eeqa
(\ref{hHpfg}) can be formally written as
\beqa
h_{n+2}&=&\frac{2p_{n+6}f^{(1)}_{n+8}}{g'_{n+12}},\n
H_{n+4}&=&\frac{2f^{(1)}_{n+8}f^{(2)}_{n+8}}{g'_{n+12}},\n
p_{n+6}&=&p_{n+6},\n
f_{n+8}&=&f^{(1)}_{n+8}-12 f^{(2)}_{n+8},\n
g_{n+12}&=&g'_{n+12} +2 f_8 H_{n+4}
\eeqa
($2 f_8 H_{n+4}$ can be absorbed in $g_{n+12}$ by redefinition). 
Therefore, provided that $2p_{n+6}f^{(1)}_{n+8}$ and $2f^{(1)}_{n+8}f^{(2)}_{n+8}$ 
are divisible by $g'_{n+12}$, they are parametrized by the four 
independent combinations
$p_{n+6}$, $f^{(1)}_{n+8}$, $f^{(2)}_{n+8}$ and $g'_{n+12}$. 
The corresponding set of $d$ and $s$ are then $3$, $2$, $2$, $0$ and 
$1,1,1,1$, respectively. Thus we have seen that, though non-simple, 
the $E_3$ bundle is  
also parametrized by the sections specified by the Casimirs 
of $A_2=SU(3)$ and $A_1=SU(2)$, which are $\{3,2\}$ and $\{2\}$, and the 
coroot expansion coefficients $-\theta=\alpha_1+\alpha_2$ and $-\theta=\alpha_1$.

For $SU(2)\times SU(2)$ bundles, the relevant curve is the one with a $D_6$ singularity.
Such a curve is realized by setting 
\beqa
 \tilde{h}_{n+2-r} &=& 0
\eeqa
in the $A_5$ curve \eqref{hHpfg}.
Consequently, $p_{n+6}=0$ in \eqref{pf1f2g}, so that the moduli space of $SU(2)\times SU(2)$ bundle 
is parametrized by $f_{n+8}^{(1)}$, $f_{n+8}^{(2)}$ and $g'_{n+12}$.
The corresponding set of $d$ and $s$ are $2$, $2$, $0$ and 
$1,1,1$, respectively. These agree with the Casimirs and the coroot expansion coefficients
of the two $SU(2)$'s.


\begin{table}[h]
\caption{ \label{}}
\centering
\small
\begin{tabular}{|c|c||c||c|c|c|}
\hline
$\begin{array}{c}\mbox{Bundle gauge}\\
\mbox{group $(=H)$}\end{array}$
& $\begin{array}{c}\mbox{Singularity}\\
\mbox{ $(=G)$ }\end{array}$& $\begin{array}{c}
\mbox{6D neutral}\\ \mbox{matter}\end{array}$ &  
$\begin{array}{c}\mbox{Independent}\\
\mbox{polynomial}\end{array}$
&$d$~& ~$s$~ \\
\hline
$E_7$
&$A_1$
&$(18n+83){\bf 1}$
&$g_{12+n}$
&0&1\\
&&&
$f_{8+n}$
&2&1\\
&&&
$g_{12+2n}$
&6&2\\
&&&
$f_{8+2n}$
&8&2\\
&&&
$h_{2n+4}$
&10&2\\
&&&
$g_{12+3n}$
&12&3
\\
&&&
$f_{8+3n}$
&14&3
\\
&&&
$g_{12+4n}$
&18&4
\\  
\hline
$E_6$
&$A_2$
&$(12n+66){\bf 1}$
&
$g_{12+n}$ 
&0&1\\
&&&
$f_{8+n}$
&2&1\\
&&&
$h_{n+2}$
&5&1\\
&&&
$g_{12+2n}$
&6&2\\
&&&
$f_{8+2n}$
&8&2\\
&&&
$H_{2n+6}$
&9&2\\
&&&
$g_{12+3n}$ 
&12&3
\\
\hline

$D_5$
&$A_3$
&$
(8n+51){\bf 1}$
&
$g_{12+n}$
&0&1\\
&&&
$f_{8+n}$
&2&1\\
&&&
$H_{n+4}$
&4&1\\
&&&
$h_{n+2}$
&5&1\\
&&&
$g_{12+2n}$
&6&2\\
&&&
$f_{8+2n}$
&8&2
\\
\hline
$A_4$
&$A_4$
&$(5n+36){\bf 1}$
&
$g_{12+n}$
&0&1\\
&&&
$f_{8+n}$
&2&1\\
&&&
$p_{n+6}$
&3&1\\
&&&
$H_{n+4}$
&4&1\\
&&&
$h_{n+2}$
&5&1\\
\hline
\end{tabular}
\end{table}

\begin{table}[h]
\small
\hskip -50ex(Cont'd)\\
\noindent
\centering
\small
\begin{tabular}{|c|c||c||c|c|c|}
\hline
%
%
%
%
%
%
~~~~~~~~~$A_3$~~~~~~~~
&~~~~~~$D_5$~~~~~~
&~$(4n+33){\bf 1}~$
&~~~~~~$g_{n+12}$~~~~~
&0&1\\
&&&
$f_{n+8}$
&2&1\\
&&&
$p_{n+6}$
&3&1\\
&&&
$H_{n+4}$
&4&1
\\
\hline
$A_2$
&$E_6$
&$(3n+28){\bf 1}$
&$g_{n+12}$
&0&1\\
&&&$f_{n+8}$
&2&1\\
&&&$p_{n+6}$
&3&1
\\
\hline
$A_1$
&$E_7$
&$(2n+21){\bf 1}$
& $g_{n+12}$
&0&1\\
&&&
$f_{n+8}$
&2&1
\\
\hline
$D_6$
&$A_1\oplus A_1$\!
&$(10n+54){\bf 1}$
&$w_{n+12}$
&0&1
\\
&&&$q_{n+8}$&2&1\\
&&&$h_{n+4}$&4&1\\
&&&$h_{n}$&6&1\\
&&&$v_{2n+12}$&6&2\\
&&&$p_{2n+8}$&8&2\\
&&&$h_{2n+4}$&10&2\\
\hline
$A_5$
&$A_2\oplus A_1$\!
&$(6n+37){\bf 1}$
&$w_{n+12}$
&0&1
\\
&&&$q_{n+8}$&2&1\\
&&&$v_{n+6}$&3&1\\
&&&$h_{n+4}$&4&1\\
&&&$h_{n+2}$&5&1\\
&&&$h_{n}$&6&1\\
\hline
$D_4$
&$D_4$
&$(6n+44){\bf 1}$
&$g_{n+12}$&0&1\\
&&&$f_{n+8}$&2&1\\
&&&$j_{n+4}$&4&1\\
&&&$k_{n+4}$&4&1\\
&&&$g_{2n+12}$&6&~2~~\\
\hline

\end{tabular}
\end{table}

\section{Why should this be so? : The Mordell-Weil lattice}
In the previous sections we have seen that the sections of a particular set of  
 line bundles coordinatizing Looijenga's weighted projective spaces can be 
automatically obtained as the coefficients of curves arising from a series of 
blow-ups in $dP_9$. They can be thought of as the four-dimensional analogue 
of the set of independent polynomials in the six-dimensional F-theory 
parameterizing the complex structure of the elliptic manifold, in which 
the gauge group of the bundle and the singularity are orthogonal to each 
other in $E_8$. 
In this section we explain why this is so.

As we stated in the previous section, the $dP_9$ we have blown up is 
supposed to be a half of a $K3$ in the stable degeneration limit, 
and the values of sections at infinity determine the spectral cover 
of the dual heterotic string theory. 

Physically, a spectral cover describes the Wilson lines in the 
elliptic fibers of the heterotic Calabi-Yau over which 
the vector bundle is defined. Therefore, if the algebra of the Wilson lines is 
$H$, the Lie algebra of the unbroken gauge subgroup $G$ is 
the commutant of $H$ in $E_8$. 
Thus it is natural to derive $H$ bundles when the singularity 
of $dP_9$ is $G$. This is a ``physical'' explanation, but 
there must also be a pure ``mathematical'' explanation which accounts 
for why the series of vector bundles are derived by the series of 
blow-ups,   
without referring to the heterotic/F-theory duality. What makes it possible 
is the structure theorem of the Mordell-Weil lattice.  

The Mordell-Weil lattice \cite{OguisoShioda} is 
the Mordell-Weil group \cite{MWtorsion} equipped with a 
certain bilinear form. The Mordell-Weil group $E(K)$ of a rational elliptic surface 
(=$dP_9$) is defined as an Abelian group of rational sections of $dP_9$, where 
$K$ is the field of rational functions of the coordinate $z$ of the base $\PP^1$ 
of $dP_9$. 
The addition of two sections is defined by the addition rule on 
an elliptic curve applied fiberwise, that is, as the addition of the two
arguments of the $\wp$ (and also $\wp'$) function parameterizing 
the two sections. As is well known, the argument variable inside $\wp$ 
(and $\wp'$) is nothing but the complex coordinate itself 
if the fiber torus is expressed as a parallelogram with the two sets of sides 
identified.

$E(K)$ is called the Mordell-Weil lattice \cite{Shioda}  
if it is endowed with a bilinear form, or a height pairing, $(P,Q)$  
for sections $P,Q\in E(K)$ such that\footnote{The fact that the arithmetic genus 
of $dP_9$ is one is taken into account here.}
\beqa
(P,Q)&=&P\cdot O+ Q\cdot O-P\cdot Q+1-\sum_{v\in{\rm singularities}} {\rm contr}_v (P,Q),\\
(P,P)&=&2+ 2P\cdot O -  \sum_{v\in{\rm singularities}} {\rm contr}_v  (P,P),
\eeqa 
where $\cdot$ denotes the intersection pairing. For each singularity $v$, 
the function ${\rm contr}_v$ of a pair of sections $P,Q$ 
is defined as  
\beqa
{\rm contr}_v (P,Q)&=&\left\{\begin{array}{ll}
0& \mbox{if $i(P)=0$ or $i(Q)=0$},\\
(C_v^{-1})_{i(P),i(Q)}& \mbox{otherwise},
\end{array}\right.
\eeqa
where $C_v$ is the Cartan matrix corresponding to the singularity $v$, and 
$i(P)$ ($i(Q)$) is either of $0,1,\ldots,{\rm rank}C_v$ 
labeling the fiber component of $v$ which (uniquely) intersects 
with the section $P$ ($Q$). The fiber labeled as ``the zeroth'' ($i=0$)  
is the one that intersects with the zero section. 

One of the remarkable results of \cite{OguisoShioda} is that 
then $E(K)$ is roughly the orthogonal complement of the singularity 
in the $E_8$ root lattice. More precisely \cite{OguisoShioda}, 
\beqa
E(K)&\simeq&L^*\otimes (T'/T),
\eeqa
where $T$ is the singularity lattice embedded into the $E_8$ root 
lattice $\Lambda_{E_8}$, $L$ is its orthogonal lattice 
with respect to 
the specified embedding into $\Lambda_{E_8}$, $L^*$ is 
the dual of $L$, and 
\beqa
T'&=&T\otimes \QQ \cap \Lambda_{E_8}.
\eeqa

This is a geometrical manifestation of the fact that  if the 
instanton is in the group $H$, the unbroken gauge group 
is the commutant of $H$ in $E_8$. By this theorem we can now 
explain why we could derive $E_N$ bundles by blowing up 
the $A_{9-N}$ singularities: As we mentioned earlier, an $E_N$ bundle 
is constructed from the spectral cover,  whose equation 
determines as the intersections with the elliptic fiber at infinity 
the Wilson lines of the vector bundle.  As one can check explicitly, 
these intersection points are extended into sections in the $dP_9$ 
\cite{CurioDonagi,DonagiWijnholt}, obtaining 
the $E_N$ weight lattice generated by the sections. The structure 
theorem of the Mordell-Weil lattice then tells us that this occurs 
precisely when the singularity lattice is the orthogonal compliment of 
the $E_N$ weight lattice, which is $A_{9-N}$.

We should mention that the rational elliptic surfaces 
with various sections and singularities are known to be identified 
\cite{SW,Lerche:1996ni,DeWolfe:1999hj,YamadaYang,FukaeYamadaYang,
EguchiSakai,MohriExceptional} 
as the total spaces of Seiberg-Witten curves for the four-, five- and 
six-dimensional so-called $E_N$ theories \cite{MorrisonSeiberg_DouglasKatzVafa}, 
where the $u$ parameter becomes the coordinate of the $\PP^1$ base.
Indeed, the curves we considered in section 2 are exactly the same as 
the ones found in \cite{Lerche:1996ni,YamadaYang}, although the 
line bundles of the sections and their relation to 
Looijenga's weighted projective spaces were not investigated there.  
We also note that the values of sections at infinity are known to 
determine the mass parameters of the gauge theory whose 
Seiberg-Witten curve (together with the $u$-plane ($\PP^1$)) is 
a rational elliptic surface allowing those sections.

The Mordell-Weil lattice also provides us with an understanding of 
the relation between the singularity and the occurrence of chiral matter 
in F-theory. (This fact was already observed and briefly mentioned 
in \cite{MizoguchiTani}.) 
In the standard explanation for the chiral matter generation \cite{KatzVafa}, 
one considers an enhanced singularity \cite{MorrisonVafa,BIKMSV}, 
at which the light membrane (in the M-theory dual picture) wrapping 
the extra shrinking two-cycle is identified as the origin of the chiral matter.
On the other hand, 
it was shown by using the Leray spectral sequence \cite{Curio,DiaconescuIonesei} 
that chiral matter is localized where one or some of the sections
of $dP_9$ goes to the zero section. Again, the relation between the 
two pictures of matter generation may also be understood
as a consequence of the structure theorem of the Mordell-Weil lattice. 
Indeed, the theorem says if some 
of the sections disappear in $dP_9$, then the singularity lattice, which 
is the orthogonal complement in $E_8$, becomes larger, 
leading to a singularity enhancement.
Also, in view of the isomorphism between 
the string junction algebra and the Picard lattice of a rational
elliptic surface \cite{FukaeYamadaYang}, it gives support to the 
understanding of matter generation in F-theory in terms of string junctions 
\cite{Tani,FFamilyUnification,MizoguchiTani}.

\section{Conclusions}
We have shown that the holomorphic vector bundles for gauge groups 
$E_N$ $(N=4,\cdots,8)$ and  $A_n$ $(n=1,2,3)$ 
can be obtained systematically by a series 
of blowing-ups in the rational elliptic surface 
according to Tate's algorithm. 
The sections of correct line bundles claimed to arise by Looijenga's  
theorem have been found automatically by this procedure. 
We have also pointed out that the sections parameterizing a Looijenga's 
weighted projective space are nothing but 
the four-dimensional analogue 
of 
the set of independent polynomials  in 
the six-dimensional F-theory  
parameterizing the complex structure of the elliptic manifold 
with a singularity orthogonal  to the gauge group of 
the vector bundle in the whole $E_8$. 
We have explained the reason for this by using the structure theorem 
of the Mordell-Weil lattice. We have also used it to elucidate 
the relation between the singularity and the occurrence of chiral matter in F-theory.  

The Mordell-Weil lattice is classified into 74 different patterns of decompositions 
of the $E_8$ root lattice, of which we have used only a few in this paper. 
It would be interesting to extend the analysis to other cases in which 
the Mordell-Weil group has a torsion \cite{MWtorsion}. 
Also, in confirming the relation between the sections and the independent 
polynomials in section 3, we have derived some examples of curves  
which possess two factorized singularities leading to an unbroken 
gauge symmetry of a direct product of two simple groups. 
A thorough investigation of this type of curves is in progress and 
will be reported elsewhere.

\section*{Acknowledgments} 
We thank H.~Itoyama and K.~Mohri for valuable discussions. 
The work of S.~M. is supported by 
Grant-in-Aid
for Scientific Research  
(C) \#25400285, 
(C) \#16K05337 
and 
(A) \#26247042
from
The Ministry of Education, Culture, Sports, Science
and Technology of Japan.


\end{document}